\documentclass[twocolumn,twocolappendix]{aastex631}

\DeclareRobustCommand{\ION}[2]{%
\relax\ifmmode
\ifx\testbx\f@series
{\mathbf{#1\,\mathsc{#2}}}\else
{\mathrm{#1\,\mathsc{#2}}}\fi
\else\textup{#1\,{\mdseries\textsc{#2}}}%
\fi}

\newcommand{\nii}         {\mbox{\rm [N{\small II}]}}
\newcommand{\oiii}        {\mbox{\rm [O{\small III}]}}

\newcommand{\Mstar}       {\mbox{$\mathrm{M_{\ast}}$}}

\newcommand{\ha}          {\mbox{H$\alpha$}}
\newcommand{\hb}          {\mbox{H$\beta$}}

\usepackage{xspace}
\usepackage{amssymb}
\usepackage{amsmath}
\usepackage{natbib}

\usepackage{hyperref}

\newcommand{\mathsc}[1]{{\normalfont\textsc{#1}}}
\usepackage{url}
\usepackage{graphicx}
\usepackage{color}
\usepackage{txfonts}

\usepackage{multirow}

\begin{document}

\title{Revisiting the Mass-Excitation (MEx) diagram using the MaNGA dataset}

\author[0000-0003-2405-7258]{J.K. Barrera-Ballesteros}
\author[0000-0001-6444-9307]{S.F.~S\'anchez}
\affiliation{Instituto de Astronom\'ia, Universidad Nacional Aut\'onoma de  M\'exico, A.P. 70-264, C.P. 04510, CdMx, M\'exico}

\begin{abstract}

The diagram comparing the flux ratio of the \oiii\ and \hb\ emission lines with the total stellar mass of galaxies (also known as the mass-excitation diagram, MEx) has been widely used to classify the ionization mechanism in high redshift galaxies, dividing star forming galaxies from those where active galactic nuclei are important. This diagram was mainly derived using single-fiber spectroscopy from the SDSS-DR7 survey. In this study, we revise this diagram using the central and integrated spectral measurement from the entire Integral Field Spectroscopic MaNGA sample. Our results suggest that along with these physical parameters, the equivalent width of the \ha\ emission line is also required to constrain the ionization mechanism of a high-redshifted galaxy. Furthermore, the location of a galaxy in the excitation-mass diagram varies depending on the use of central or integrated properties. 

\end{abstract}

\section{Introduction} \label{sec:intro}

Characterizing the ionization mechanisms in galaxies is crucial to understand their formation and evolution. Depending on whether the emission from the ionized gas of a galaxy is dominated by formation of stars, or by an active galactic nucleus, different physical processes could determine the fate of the galaxy. This is significantly evident for high-redshift galaxies as one or the other mechanism could result in different properties of a galaxy at the present time.  

One of the most useful tools to characterize the ionization mechanism of a galaxy is the emission-line diagnostic diagrams usually derived in the optical. These diagrams compare the flux ratio from two pairs of emission lines. Depending on the location of a galaxy in one or several of these diagrams it can be classified either as star-forming (SF) or as active-galactic nucleus (AGN). From these diagrams, the most well-known is the so-called BPT diagram that compares the \oiii/\hb\ against the \nii/\ha\ emission line ratios \citep{BPT_1981}. Given the observational constraints, the BPT diagram has been extensively used mostly for nearby galaxies \citep[$z \sim$ 0.1]{Kauffmann_2003}. Given the fact that the emission from \nii, and \ha\ lines for galaxies at high-$z$ galaxies is not observable in the optical, it is not possible to use the BPT diagram to characterize the ionization mechanism for those galaxies. To mitigate this, \citet{Juneau_2014} derive an alternative diagram that makes use only of the \oiii/\hb\ lines ratio and the total stellar mass, \Mstar, also known as the Mass-Excitation (MEx) diagram. Taking advantage of the large spectroscopic sample provided by the SDSS-DR7 survey, the authors derived demarcation lines in the MEx diagram between star-forming and AGN galaxies.  Although the MEx diagram and its demarcation lines have been extensively used in order to characterize the main ionization mechanism from high-$z$ galaxies, it was derived using only central measurements of galaxies provided by the single-fiber spectroscopy from the SDSS-DR7 dataset. This could hamper the classifications from high-$z$ galaxies, as usually the spectral measurements from these galaxies are integrated measurements. Furthermore, the MEx diagram is based only in the emission line ratios used in optical diagnostic diagrams. However as explored recently by \citet[][and references therein]{Sanchez_2020ARAA}, the ionization stage of a galaxy/region in the BPT diagram is defined by both, its location in that diagram and the equivalent width of the \ha\ emission line EW(\ha). In this study we report the MEx using the largest sample of galaxies observed with Integral Field Spectroscopy (IFS): The SDSS-IV MaNGA survey \citep{Bundy_2015}. The MaNGA survey has observed more than 10000 galaxies in the nearby universe, which allows us to explore the difference between the MEx derived using central properties \citep[a similar procedure as the one using single-fiber spectroscopy by ][]{Juneau_2014}, and integrated properties (the measurements derived from high-$z$ galaxies). 

\section{Results} 
\label{sec:res}

\begin{figure*}[htb]
    \begin{minipage}[t]{.5\textwidth}
        \centering
        \includegraphics[width=\textwidth]{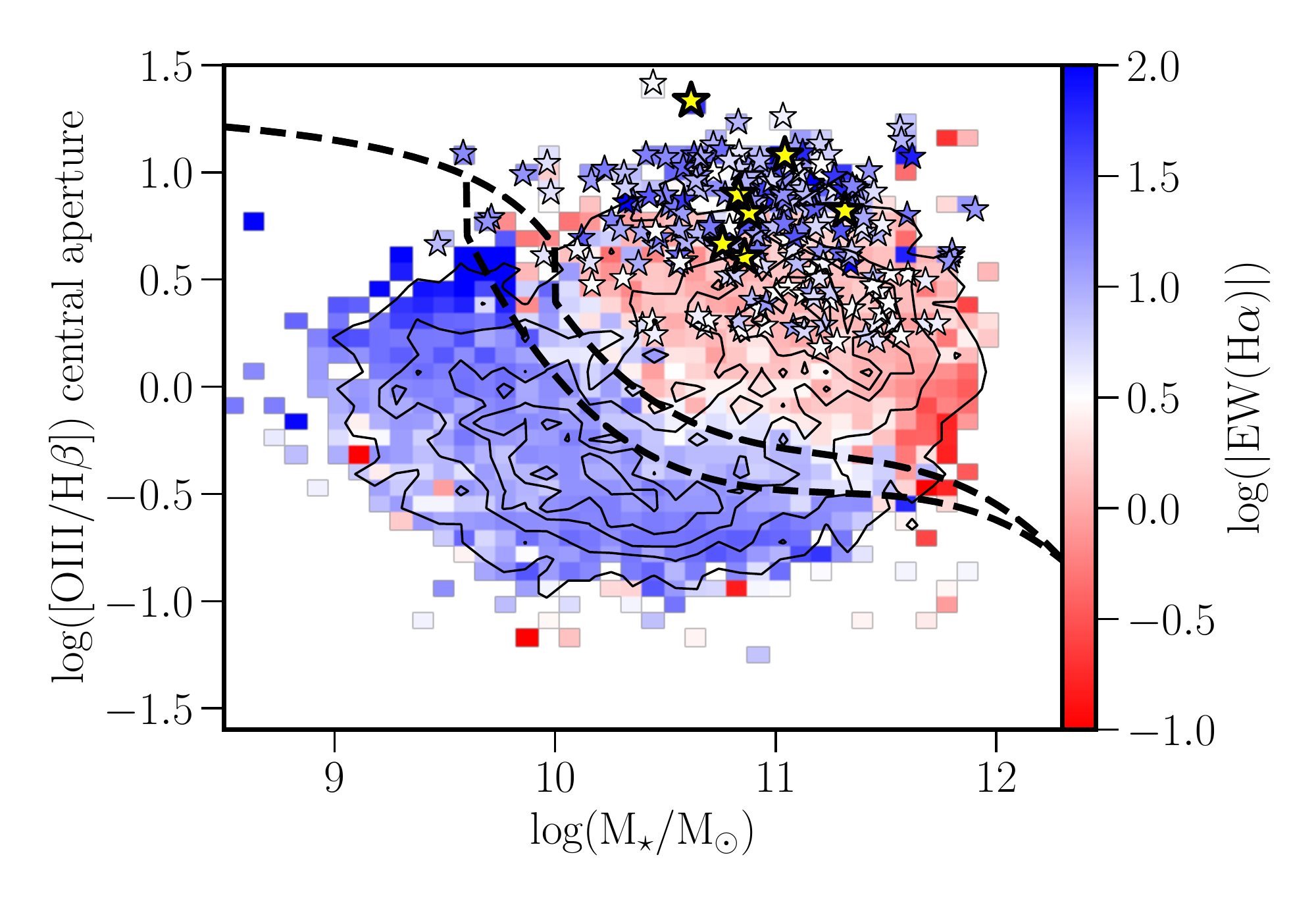}
    \end{minipage}
    \hfill
    \begin{minipage}[t]{.5\textwidth}
        \centering
        \includegraphics[width=\textwidth]{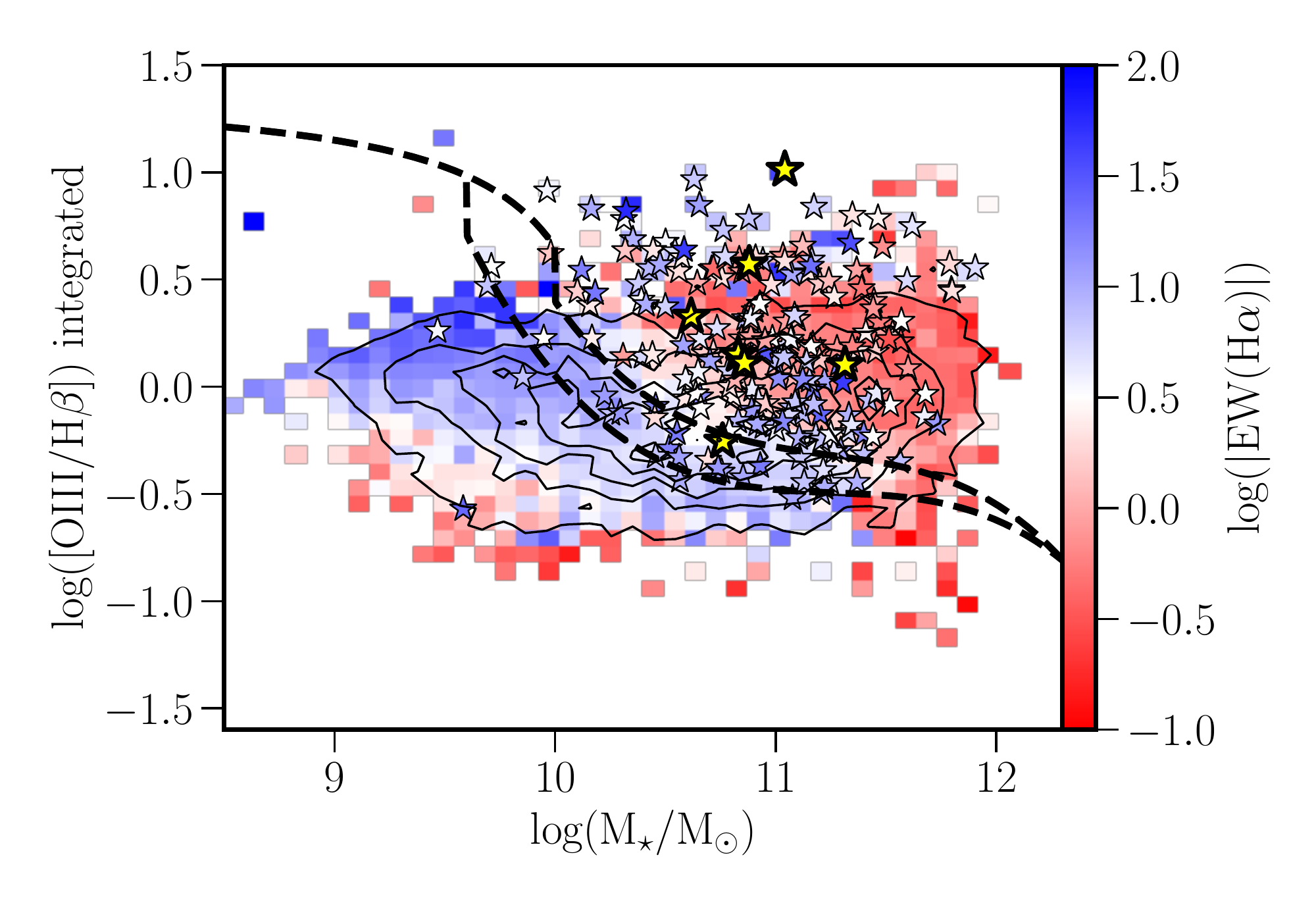}
    \end{minipage}  
    \caption{The distribution of the entire MaNGA sample in the mass-excitation diagram (MEx) using central (left panel) and integrated (right panel) measurements. In both panels the distribution is binned and color-coded according to the average EW(\ha) within each bin. Yellow and color-codded stars represent the AGN Seyfert 1 and 2 samples selected by \cite{Sanchez_2022}, respectively. In both panels the dashed lines represent the demarcation lines derived by \citet{Juneau_2014}.}
    \label{fig:MEx}
\end{figure*}

The dataset used in this study has been taken from the publicly available measurements reported in \citet{Sanchez_2022}. In that study, we reported the central (2.5 arcsec aperture) and integrated measurement for the entire MaNGA sample. The reader is referred to \citet[][and references therein]{Sanchez_2022} to learn more about the details of the observation such as, sample selection, observation strategy, data reduction, and data analysis. In Fig.~\ref{fig:MEx} we show the binned distribution of the MaNGA sample in the MEx diagram for the aforementioned spatial scales: central and integrated measurements (left and right panels, respectively). In both panels we color-coded each bin of the distribution according to its average EW(\ha). As reference, we include the demarcation lines proposed by \citet[][dashed lines]{Juneau_2014}. Above or below these lines, \citet{Juneau_2014} classified galaxies as AGN or star-forming, respectively. Furthermore, as a practical case of use of the integrated measurements from the MaNGA dataset, in \citep{Sanchez_2022} we select those galaxies classified as AGN Seyfert 1 and 2 type (yellow and color-coded stars in Fig.~\ref{fig:MEx}). For central measurements, we find a clear separation between galaxies with high and low EW(\ha) in the MEx diagram. Furthermore, the demarcation lines proposed by \citet{Juneau_2014} describe well this segregation with galaxies with low and high  EW(\ha) lie above and below the demarcation lines, respectively. This suggest that for the majority of the population of galaxies, the \citet{Juneau_2014} demarcation lines separate star-forming from non star-forming galaxies. However most of the well-characterized AGN galaxies lie above the demarcation lines (except for one galaxy lying below those lines). Furthermore, the color code from the Seyfert 1 galaxies indicates that the EW(\ha) is significantly larger than those galaxies with similar position in the MEx diagram. This is clear evidence that although AGN lie above the demarcation lines, it is required the EW(\ha) as an additional parameter to confidently classify a galaxy as AGN in the MEx diagram.

On the other hand, when we plot the integrated properties we find a different distribution of the galaxies in the MEx diagram. In contrast to central measurements, the separation between galaxies with high and low EW(\ha) is not longer very evident. There is a loose segregation of galaxies with respect to \Mstar: low-mass and massive galaxies tend to have low and high EW(\ha), respectively. Furthermore, we note a rather flat/narrower distribution of the  \oiii/\hb\ ratio in comparison to central measurements. The demarcation lines using single fiber spectroscopy are no longer clearly segregated galaxies with low EW(\ha) from those with high EW(\ha). As expected from the negative radial distribution of the emission line ratios \citep[e.g.,][]{Barrera-Ballesteros_2022}, we find that the \oiii/\hb\ ratio drops from central to integrated measurements in the well-selected AGN galaxies as well as their EW(\ha). As a consequence, a significant fraction of Seyfert 1 galaxies lie  between the two demarcation lines proposed by \citet{Juneau_2014}.

In conclusion, using the MaNGA survey -- the largest sample of galaxies in the nearby universe  with IFS information -- we find that the MEx diagram and the demarcation lines derived by \citet{Juneau_2014} is a good tool to segregated star-forming from AGN galaxies. However, our results also show that in order to confidently classify a galaxy as an AGN it is required to measure the central emission as well as to have an estimation of the EW(\ha). Otherwise, integrated measurement with no information of EW(\ha) could lead to select a non-SF galaxy where the ionization mechanism is not an AGN but any other mechanism such as shock or heating from old hot stars. 


\begin{acknowledgments}
J.B-B acknowledges support from the grant IA-101522 (DGAPA-PAPIIT, UNAM) and funding from the CONACYT grant CF19-39578. SFS and J.B-B are grateful for the support of the CONACYT grant CB-285080 and FC-2016-01-1916, and funding from the PAPIIT-DGAPA-IN100519 (UNAM) projects.
\end{acknowledgments}

%






\bibliography{sample631}{}
\bibliographystyle{aasjournal}



\end{document}